\documentclass[prb,twocolumn,floats,showpacs,superscriptaddress]{revtex4}

\usepackage{bm}
\usepackage{graphicx}% Include figure files
\usepackage{amssymb}
\usepackage{amsfonts}
\usepackage{amsmath}
\usepackage{epstopdf}
\usepackage{color}

\def \IFPAN{Institute of Physics, Polish Academy of Sciences, al. Lotnik\'{o}w 32/46, 02-668 Warsaw, Poland}
\def \UB{Institute of Physics, Kazimierz Wielki University, Powstanc\'{o}w Wielkopolskich 2, 85-064 Bydgoszcz, Poland}

\begin{document}

\title{Metastability of Mn$^{3+}$ in ZnO driven by strong $d$(Mn) intrashell Coulomb repulsion: experiment and theory}

\author{A. Ciechan}\affiliation{\IFPAN}
\author{H. Przybyli\'nska}\affiliation{\IFPAN}
\author{P. Bogus\l awski}\email{bogus@ifpan.edu.pl}\affiliation{\IFPAN}\affiliation{\UB}
\author{A. Suchocki}\affiliation{\IFPAN}\affiliation{\UB}
\author{A. Grochot}\affiliation{\IFPAN}
\author{A. Mycielski}\affiliation{\IFPAN}
\author{P. Skupi\'nski}\affiliation{\IFPAN}
\author{K. Grasza}\affiliation{\IFPAN}
 
\date{\today}

\begin{abstract}
Depopulation of the Mn$^{2+}$ state in ZnO:Mn upon illumination, monitored by quenching of the Mn$^{2+}$ EPR signal intensity, was observed at temperatures below 80~K. Mn$^{2+}$ photoquenching is shown to result from the Mn$^{2+}$ $\to$ Mn$^{3+}$ ionization transition, promoting one electron to the conduction band. Temperature dependence of this process indicates the existence of an energy barrier for electron recapture of the order of 1~meV. 
GGA$+U$ calculations show that after ionization of Mn$^{2+}$ a moderate breathing lattice relaxation in the 3+ charge state occurs, which increases energies of $d$(Mn) levels. 
At its equilibrium atomic configuration, Mn$^{3+}$ is metastable since the direct capture of photo-electron is not possible. 
The metastability is mainly driven by the strong intra-shell Coulomb repulsion between $d$(Mn) electrons. Both the estimated barrier for electron capture and the photoionization energy are in good agreement with the experimental values.
\end{abstract}

\pacs{71.55.-i, 71.55.Gs, 76.30.-v, 71.15.Mb}
% 76.30.-v  Electron paramagnetic resonance and relaxation
% 71.15.Mb  Density functional theory, local density approximation, gradient and other corrections
% 71.55.-i  Impurity and defect levels
% 71.55.Gs  II-VI semiconductors

\maketitle

%%%%%%%%%%%%%%%%%%%%%%%%%%%%%%%%%%%%%%%%%%%%%
\section{\label{sec1}Introduction}
ZnO is a promising material for photocatalysis~\cite{Maeda} and photovoltaic applications.~\cite{Law, Riaz} Mn substituting for the divalent cation in ZnO introduces a Mn$^{2+}$/Mn$^{3+}$ level located in the forbidden gap.~\cite{Johnson} The mid-gap position of Mn$^{2+}$ has been already practically utilized and powers the research on water splitting.~\cite{Maeda} 
Mn-doped ZnO also exhibits a chromatographic effect: the undoped transparent crystals upon doping with Mn turn reddish-brown due to the strong absorption interpreted as Mn$^{2+}\to$ (Mn$^{3+}, e_{CB}$) photo-ionization transition,~\cite{Johnson} where $e_{CB}$ denotes a photoelectron in the conduction band. This absorption is accompanied by photoconductivity.~\cite{Johnson}  
The nature of this transition has been inferred only indirectly. Though the presence of Mn in the 2+ charge state in ZnO was detected with use of electron paramagnetic resonance (EPR),~\cite{Hausmann, Chikoidze} no optical spectra related to 
intra-center transitions of Mn$^{2+}$ were observed. Since these transitions can occur at energies higher than the observed photo-ionization band, it was concluded that the excited states of Mn$^{2+}$ are degenerate with the conduction band of ZnO~\cite{Godlewski2010}, consistent with the midgap position of the Mn$^{2+}$ energy level. However, no direct evidence of the depopulation of the Mn$^{2+}$ state under illumination was presented so far.

In this paper we study directly the occupancy of Mn$^{2+}$ ions under illumination by means of photo-EPR spectroscopy. We observe a temperature dependent decrease of the EPR signal intensity under excitation with light of energies corresponding to the Mn related absorption band. The kinetics of the EPR signal photo-quenching points out to a process involving photocarriers and the Mn ions directly.
First principles calculations indicate that the observed photoquenching is due to a transition of Mn$^{3+}$ to a metastable state after photoionization. Metastability of defects and/or dopants typically originates in strong lattice relaxations after the change of the defect charge state. 
In the case of the As antisite in GaAs studied in the past (the EL2 center), optical excitation is followed by a large displacement, exceeding 1~\AA, of the defect towards the metastable interstitial site.~\cite{EL2, EL2_prl} A similar mechanism is operative also in the case of donors, which can acquire the DX configuration when a shallow donor captures an electron and becomes a deep one with a strongly localized electronic state in the band gap,~\cite{Chadi, Dobaczewski, BBdx, Wetzel, Thio} and in the case of native defects,~\cite{Lany_DX} where the (meta)stability is responsible for quenching of doping efficiency. 
A metastable configuration can also consist in a breathing-like displacement of the surrounding host atoms.~\cite{Jones, Schmidt} According to the present results, metastability of Mn in ZnO also requires substantial lattice relaxations induced by the change of the charge state. However, a novel factor that drives metastability of Mn$^{3+}$ is the strong intracenter Coulomb coupling between the $d$(Mn) states, which prevents the electron capture by Mn$^{3+}$ followed by recombination. Finally, regarding the absorption measurements, the calculations predict that intracenter transitions should occur at energies higher than photoionization, in agreement with experimental data. 

The paper is organized in the following way: in Sec.~\ref{sec2} the experimental setup and results are presented and discussed. 
In Sec.~\ref{sec3}, details of the theoretical approach, based on the Generalized Gradient Approximation (GGA) to the Density Functional Theory, are given. 
The $+U$ corrections~\cite{Anisimov1991, Anisimov1993, Cococcioni} 
are applied to $d$(Zn), $p$(O), and the $d$(Mn) shell. 
The proposed mechanism of metastability of the photoionized Mn is presented in Sec. \ref{sec3} D. 
Section~\ref{sec4} summarizes the obtained results. 

%%%%%%%%%%%%%%%%%%%%%%%%%%%%%%%%%%%%%%%%%%%%%
\section{\label{sec2}Results and discussion}
\subsection{\label{sec2a}Experimental methods}
Mn doped ZnO single crystals were grown by chemical vapor transport.~\cite{mycio}  For the photo-EPR experiments the Mn concentration of 0.2~\% was chosen, as it ensures well resolved, narrow-line EPR spectra of Mn$^{2+}$. The sample was placed in an Oxford Instruments He gas flow cryostat enabling temperature dependent measurements in the range of 3-300~K. The EPR experiments were performed at 9.5~GHz, with use of a BRUKER ESP300 spectrometer 
equipped with Oxford Instruments ESR 900 cryostat operating in the temperature range 1.8-300~K. 
The magnetic field was oriented perpendicular to the $c$-axis of the crystal. The sample was illuminated at right angle to the magnetic field direction with a set of laser diodes of wavelengths varying from 445~nm to 980~nm. For power dependent measurements a set of gray filters was employed. 

\subsection{\label{sec2b}Experimental results} 
The as-grown ZnO:Mn 0.2~\% sample is highly resistive, in contrast to the n-type conductivity of undoped ZnO crystals grown with the same method.  A part of the Mn ions occurs in the Mn$^{2+}$ charge state and can be easily detected by EPR. Annealing the crystal in hydrogen atmosphere leads to a substantial (more than fivefold) increase of the Mn$^{2+}$ ion fraction accompanied by the appearance of n-type conductivity.

\begin{figure*}[th!]
\begin{center}
\includegraphics[width=8.3cm]{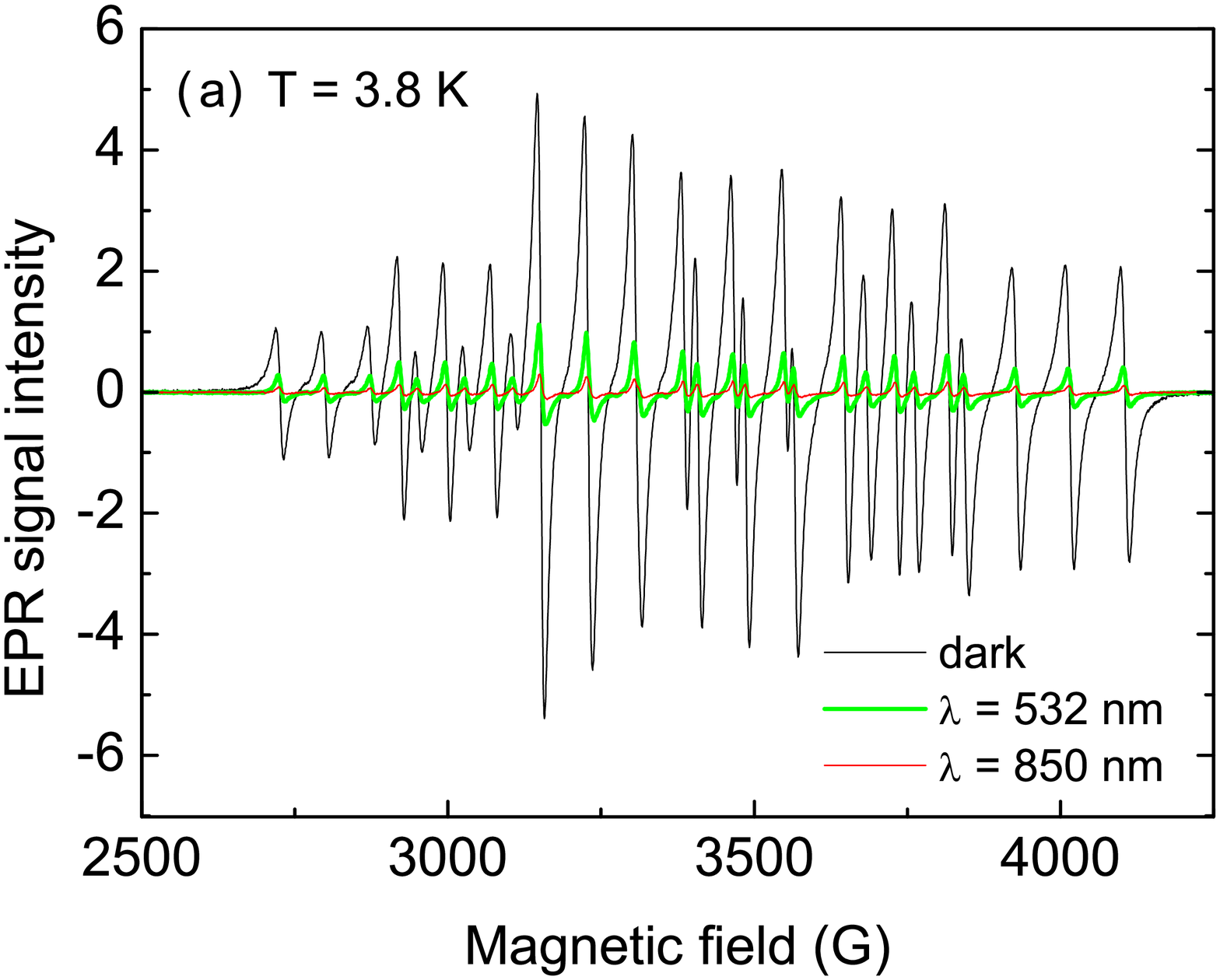}
\includegraphics[width=8.3cm]{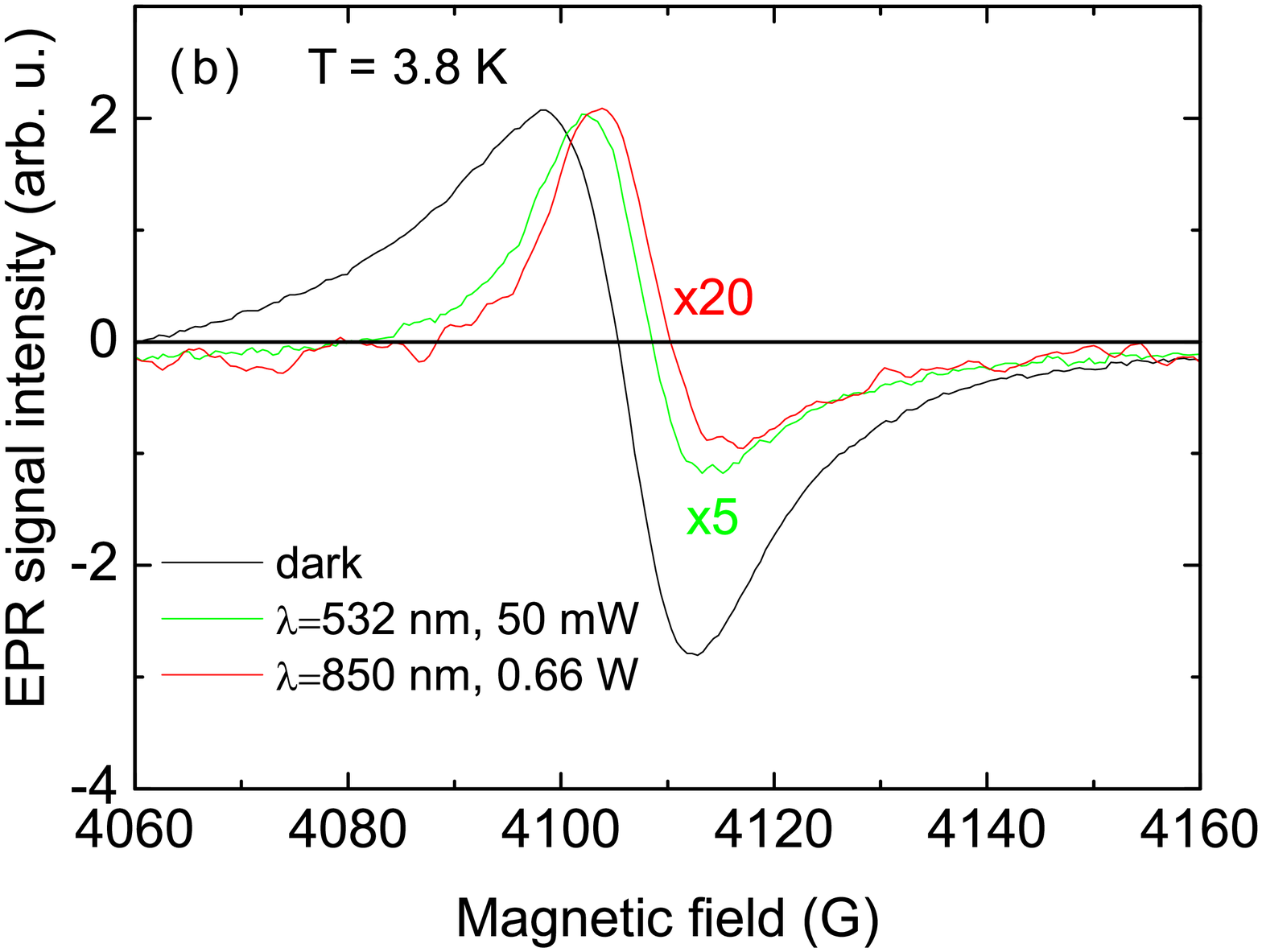}
\caption{\label{fig1} 
(a) EPR spectrum of Mn$^{2+}$ in ZnO with $B$ perpendicular to the 
$c$-axis. Black line shows the signal in the dark, green and red traces are recorded under illumination with 532~nm (50~mW)  and 850~nm (660~mW) laser lines, respectively. (b) EPR signal of (a) in the magnetic field range 4060-4160~G showing the change of the lineshape and the shift of the line position under illumination.
}
\end{center}
\end{figure*}
Figure~\ref{fig1}a shows the EPR spectrum of Mn$^{2+}$ in the as grown ZnO:Mn sample at 3.8~K taken with the magnetic field oriented perpendicular to the $c$-axis of the crystal. The spectrum consists of 30 partly overlapped resonances grouped into 5 sextets. The five so-called fine structure groups stem from allowed $\Delta M_S=\pm 1$ transitions between electronic spin levels of a $d^5$ ion with the electronic spin of $S=5/2$. Each group consist of 6 equally intense lines due to hyperfine interaction with the $I=5/2$ nuclear spin of Mn$^{55}$. The spectrum is characteristic of isolated Mn$^{2+}$ ions in ZnO.~\cite{Hausmann, Chikoidze} Analysis of the angular dependence of the resonance peak positions measured~\cite{note2} yields the spin Hamiltonian parameters $g = 2.0025 \pm 0.0003$, $D = -248.53 \pm 0.07$~G, $A_\parallel = 80.4 \pm 0.1$~G, $A_\perp = 80.3 \pm 0.3$~G, and $a = 4.0 \pm 0.2$~G at 3~K, consistent with earlier studies.~\cite{Johnson, Gluba} 

Apart from the EPR spectrum of isolated Mn$^{2+}$ ions, no other EPR signals were detected in our crystals, in particular neither complexes of Mn$^{2+}$ with other defects (up to second nearest neighbors), nor spectra related to Mn-Mn pairs were observed. 

Illumination with light in the 980 - 445~nm range leads to a drastic reduction of the detected EPR signal intensity of Mn$^{2+}$. Exemplary spectra recorded at 3.8~K under illumination with 532~nm and 850~nm laser lines are shown in Fig.~\ref{fig1}b. The laser power was 50 and 660~mW, respectively. Not all of the observed signal reduction can be attributed to a change of Mn$^{2+}$ concentration alone. The dominant mechanism of the EPR intensity quenching shown in Fig.~\ref{fig1} comes from the skin effect, {\it i.e.}, absorption by photogenerated free carriers, which reduces the microwave penetration depth and hence the effective volume of the sample. The skin effect manifests itself in a change of the resonance line shape from Gaussian to Dysonian, as shown in Fig.~\ref{fig1}b. In addition, we observe a small shift of the resonance line positions towards higher magnetic fields under illumination. 
This shift is due to exchange interaction between localized magnetic moments of 
  Mn$^{2+}$ and free carrier spins,~\cite{StoryPRL} an analogue of the Knight shift in nuclear magnetic resonance. Both the change of the EPR lineshape and the shift of the resonance fields directly prove that illumination with light in the whole wavelength range (445 - 980~nm) studied leads to generation of free carriers. 

To eliminate the skin effect the sample was thinned down to 100~$\mu$m. This thickness was found to be sufficient to ensure microwave penetration of the entire sample. We no longer observed changes of the line shape accompanying the reduction of the Mn$^{2+}$ EPR signal intensity upon illumination. We can also exclude another possible source of intensity decrease in our experiment, {\it i.e.}, sample heating due to incident laser power. Since the fine structure $-5/2\to -3/2$ (high field resonances in Fig.~\ref{fig1}a) and $3/2\to 5/2$ transitions (low field resonances) have the same probability, the difference in the intensities of the high field and low field resonance lines reflects the difference in the thermal population of the -5/2 and 3/2 levels. At low temperatures (see Fig.~\ref{fig1}a) the intensity of high field resonances is more than twice higher than that of the low field ones. 
With increasing sample temperature the intensity ratio decreases, and at 300~K both EPR resonances are almost equally intense. Even under illumination with 2.4~W at the lowest applied wavelength of 980~nm we observed no measurable change in the EPR signal intensity ratio between the $-5/2\to -3/2$ and $3/2\to 5/2$ resonances. 
Thus, any light induced changes of the EPR signal intensity measured in the so prepared sample reflect solely the change in the occupancy of the manganese 2+ charge state. Unless explicitly specified, all further data reported here refer to measurements performed on the thin sample.

\begin{figure*}[t!]
\begin{center}
\includegraphics[width=8.3cm]{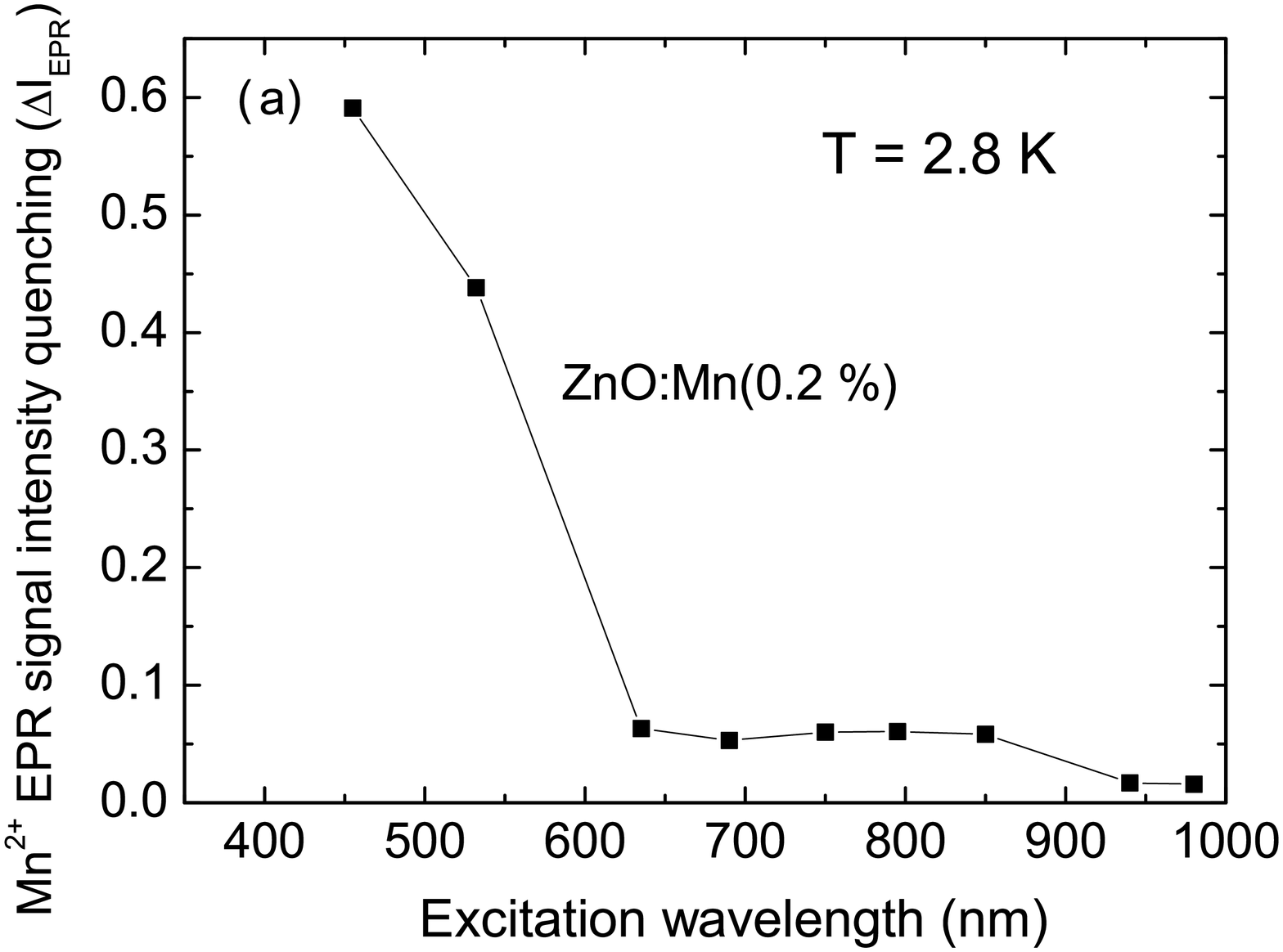}
\includegraphics[width=8.3cm]{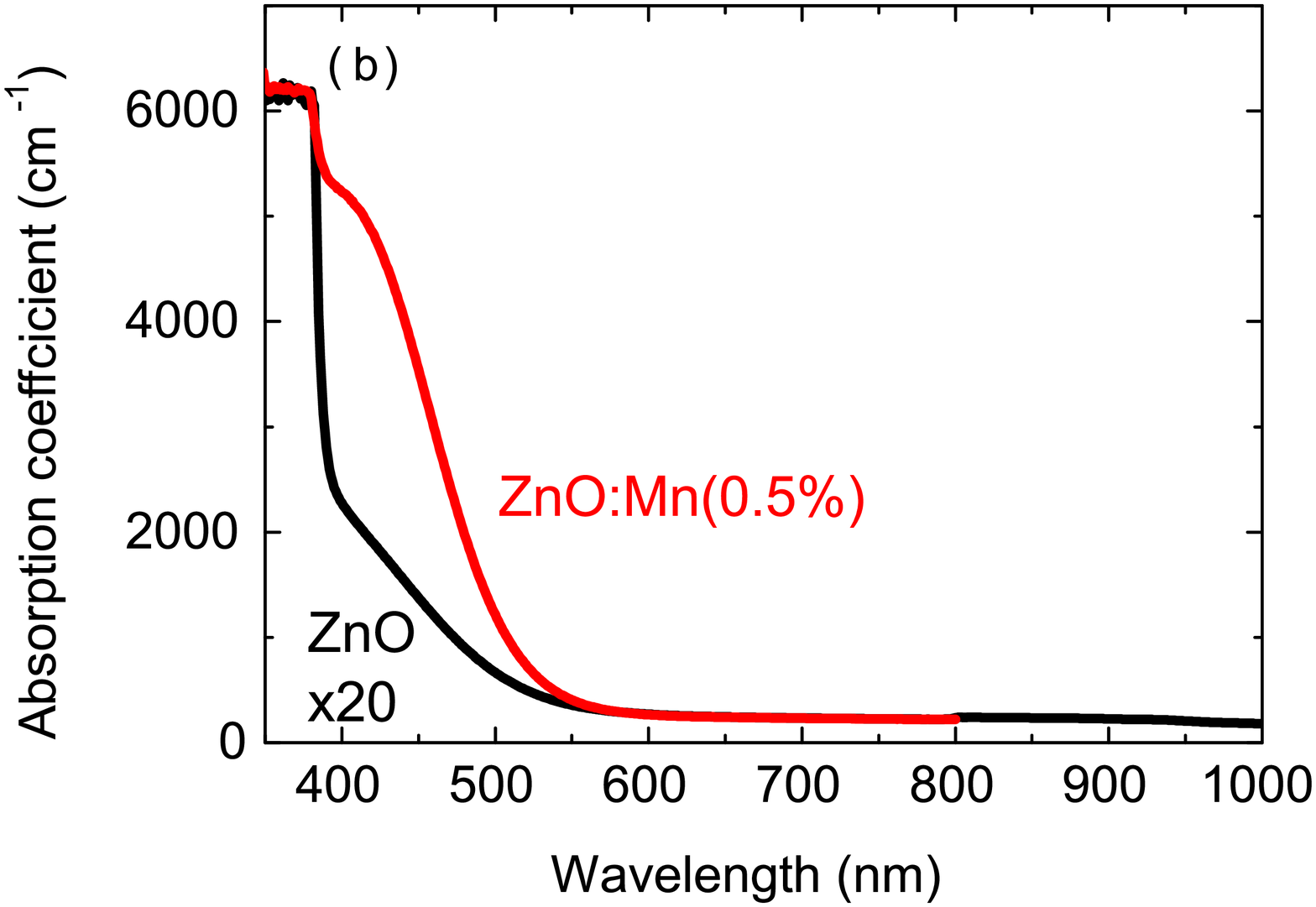}
\caption{\label{fig2} 
(a) Quenching of Mn$^{2+}$ EPR intensity, $\Delta I_{EPR}$, at T=2.8~K as a function of excitation wavelength measured in the thin sample. (b) Spectral dependence of the absorption coefficient for both pure ZnO and ZnO doped with 0.5~\% Mn at 300~K.
}
\end{center}
\end{figure*}
The spectral dependence of the Mn$^{2+}$ EPR signal photoquenching is presented in Fig.~\ref{fig2}a. Depicted is the relative reduction of the EPR signal intensity, $\Delta I_{EPR}$, under illumination at a constant power of 50~mW. $\Delta I_{EPR}$ is defined as the difference between the signal intensities in the dark, $I_{EPR}(dark)$, and under illumination, $I_{EPR}(ilumin)$, divided by the dark intensity: 
\begin{equation}
\label{eq1} 
\Delta I_{EPR}=\frac{I_{EPR}(dark)-I_{EPR}(ilumin)}{I_{EPR}(dark)}.
\end{equation}
For comparison, the room temperature absorption spectra of ZnO:Mn 0.5~\% and undoped ZnO are shown (Fig.~\ref{fig2}b). As can be seen in Fig.~\ref{fig2}b, the onset of the Mn-related absorption is close to 620~nm (2~eV), which is consistent with the optical ($E_{opt} = 2.6 \pm 0.1$~eV) and thermal ($E_{th} = 2.1 \pm 0.1$~eV) ionization energies determined in Ref.~\onlinecite{Godlewski2009} for the postulated Mn 2+ to 3+ photoionization transition.  The agreement between the spectral dependence of the Mn$^{2+}$ EPR signal photoquenching below 600~nm in Fig.~\ref{fig2}a and the absorption shown in 
Fig~\ref{fig2}b proves unambiguously that the absorption band is indeed due to photoionization of Mn$^{2+}$. 
However, above 600~nm there is a non-vanishing tail in $\Delta I_{EPR}$, which we attribute to an indirect quenching mechanism, {\it i.e.}, capture of holes generated in the photoneutralization processes of other defects present in the sample. Although the tail seems to be weak at the excitation power of 50~mW, at high incident powers $\Delta I_{EPR}$ due to the indirect mechanism is comparable to that observed for direct photoionization of Mn$^{2+}$. This means that the concentration of the defects involved is not negligible. 

We note that the spectral dependence of $\Delta I_{EPR}$ in Fig.~\ref{fig2} may not be very accurate, as in the photoquenching experiment different light sources for each wavelength were used. Although the light path was always optimized for the maximum response, there still remains the error connected to the difference in the spot sizes of the laser diodes. 

\begin{figure*}[t!]
\begin{center}
\includegraphics[width=8.3cm]{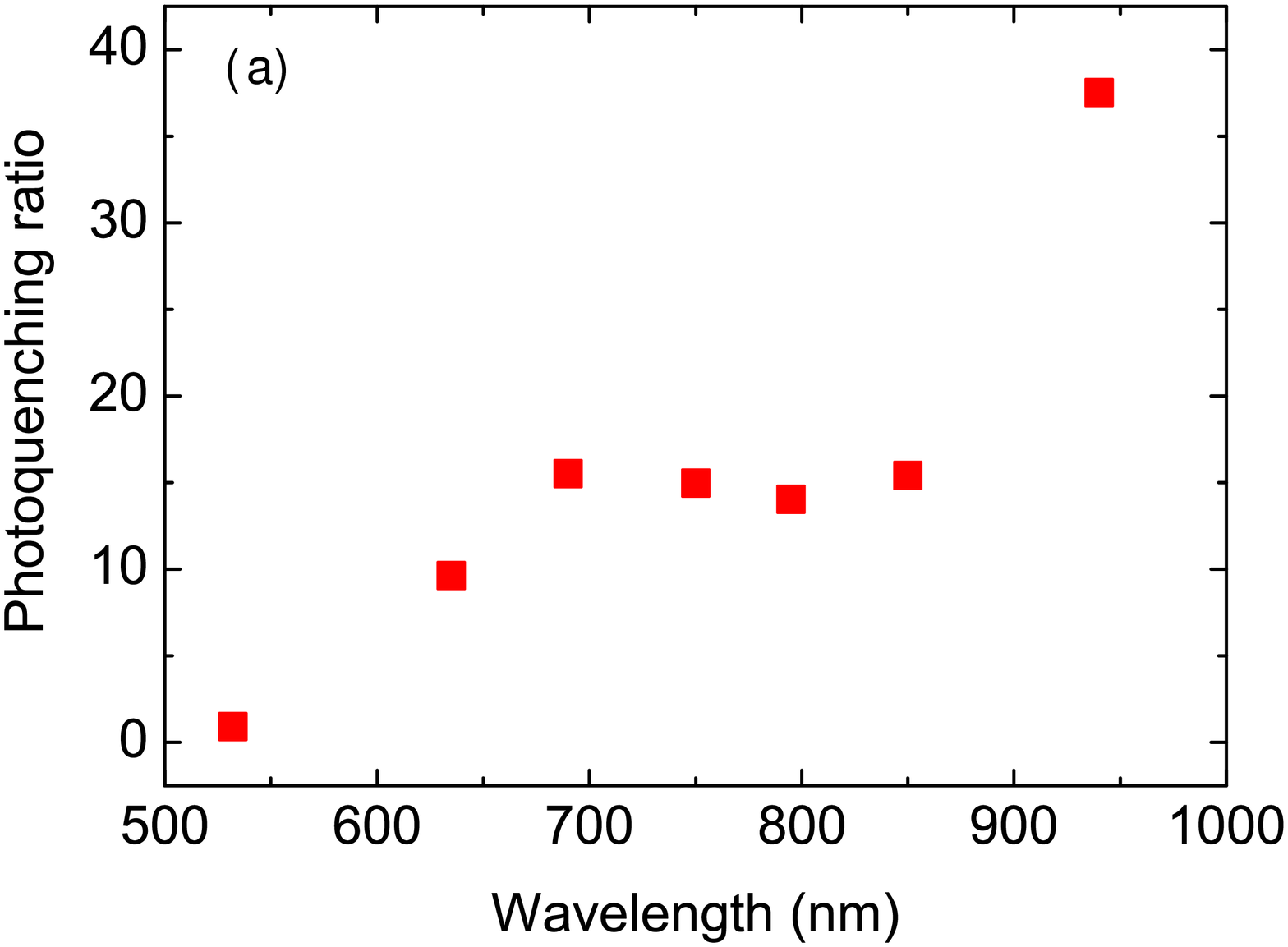}
\includegraphics[width=8.3cm]{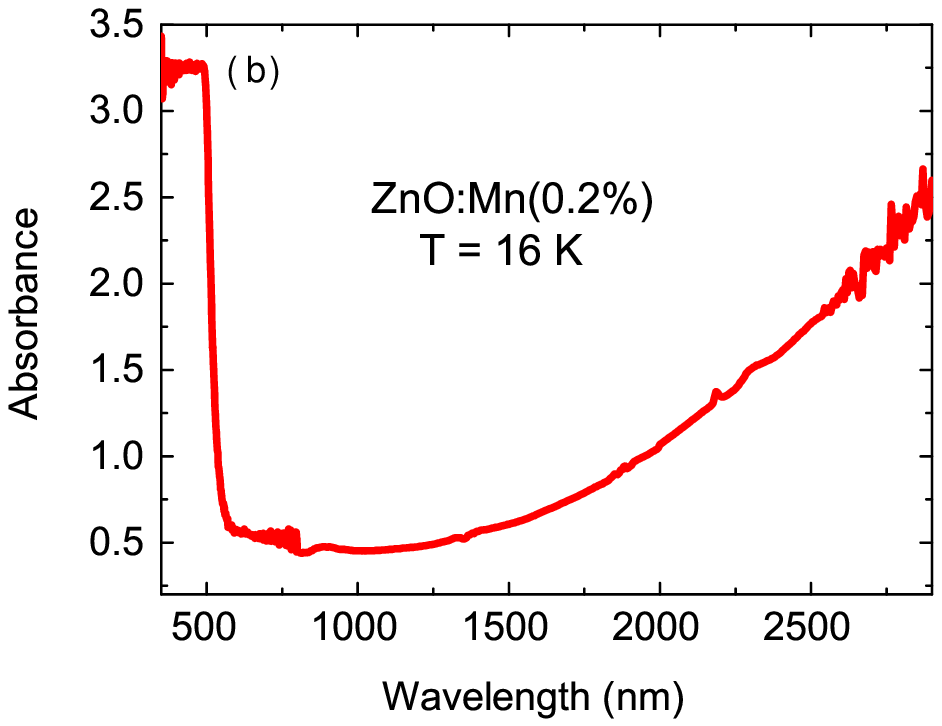}
\caption{\label{fig3} 
(a) Ratio $\Delta I_{EPR}$(thick)/$\Delta I_{EPR}$(thin) in thick and thin samples vs. excitation wavelength. (b) Spectral dependence of absorbance at $T$=16~K. Note the difference wavelength window in (a) and (b).
}
\end{center}
\end{figure*}
As already mentioned, in the thick ZnO:Mn sample we observe an additional reduction of the Mn$^{2+}$ signal intensity under illumination, related to microwave absorption by free carriers. This reduction should increase with the concentration of photogenerated carriers as the effective volume penetrated by microwaves decreases. If the free carriers would originate solely from Mn$^{2+}$, the spectral dependencies measured in thin and thick samples should scale with the behavior of the Mn$^{2+}$ photoionization band. In Fig.~\ref{fig3}a we show the spectrally dependent quenching of the EPR signal intensity of Mn$^{2+}$ measured in the thick sample  divided by the quenching measured after the sample was thinned down to 100~$\mu$m, $\Delta I_{EPR}$(thick)/$\Delta I_{EPR}$(thin). The measurements were performed at a constant temperature of 3.8~K. 
As can be seen, the photoquenching ratio increases with increasing wavelength, in contrast to the behavior expected if  Mn$^{2+}$ photoionization would be the only mechanism of free carrier generation. 
This result demonstrates that there is at least a second channel of carrier photogeneration, dominant for wavelengths longer than 500~nm.  Absorption extending to even longer wavelengths than applied in the photoquenching experiment is also observed in the optical spectrum of ZnO:Mn 0.2~\% measured at low temperatures (16~K) shown in Fig.~\ref{fig3}b. The nature and number of the defects responsible for carrier generation cannot be determined in our experiment as they give no paramagnetic signal. In the studied samples no EPR signal other than that of Mn$^{2+}$ was detected.

\begin{figure}[t!]
\begin{center}
\includegraphics[width=8.3cm]{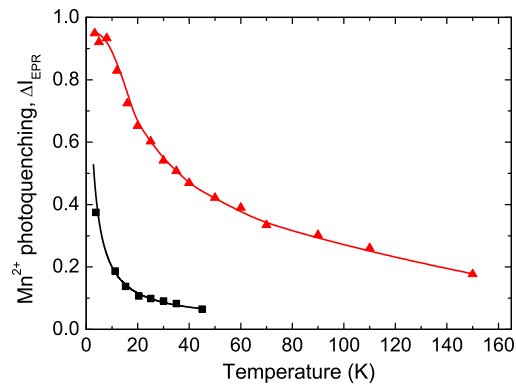}
\caption{\label{fig4} 
Temperature dependencies of Mn$^{2+}$ EPR signal photoquenching ($\Delta I_{EPR}$) in the as-grown thin sample (squares) and in the same sample after hydrogenation (triangles) illuminated with 532~nm light at 50~mW. The solid lines were calculated assuming a thermally activated recapture process.
}
\end{center}
\end{figure}
In contrast to absorption measurements, where the photoionization transition is observed up to room temperature, in the photo-EPR experiment photoquenching was found to be temperature dependent, as shown in Fig.~\ref{fig4}. This is due to a fundamental difference between the two experiments. Whereas in absorption the signal is predominantly proportional to the occupancy of the initial state (Mn$^{2+}$), in photo-EPR the change of the signal is proportional to the transient occupancy of the final state (Mn$^{3+} + e_{CB}$). This means that a fast recapture of the ionized electron by Mn$^{3+}$ decreases photoquenching. In other words, photoquenching can only be observed if the occupancy of the final state is metastable. This can be achieved in two ways: either the photoionized electrons are trapped on other defect centers, or the recapture proceeds via an energy barrier. In both processes the recapture has a thermally activated character. 
Shown in Fig.~\ref{fig4} are the temperature dependencies of $\Delta I_{EPR}$ in the as-grown sample (squares) and the same sample after hydrogenation (triangles) illuminated with 532~nm light at 50~mW. 
The solid lines were calculated with use of the following simple relation:
\begin{equation}
\label{eq2} 
(1-\Delta I_{EPR})^{-1}= A[1+B\exp(E_B/kT)], 
\end{equation}
where  the parameter $A$ is the ratio of  the concentration of Mn$^{2+}$ ions in the dark to the total 
Mn concentration. 
Parameter $B=I\sigma/(nr)$ depends on the electron concentration $n$, light intensity $I$, Mn$^{2+}$ 
absorption cross section $\sigma$, and the temperature independent part of electron capture probability by Mn$^{3+}$, $r$. $E_B$ is the energy barrier for recapture of photoionized electrons. This relation was obtained assuming that the concentration of photogenerated electrons is much higher than that photoionized from Mn$^{2+}$. Unfortunately, as we do not know the fraction of occupied Mn$^{2+}$ ions in the sample 
(given by the parameter $A$), the parameters cannot be determined independently of each other. 
However, we obtain a reasonable agreement with experimental data with a finite barrier with the lowest estimated value of ~meV.

\begin{figure}[t!]
\begin{center}
\includegraphics[width=8.3cm]{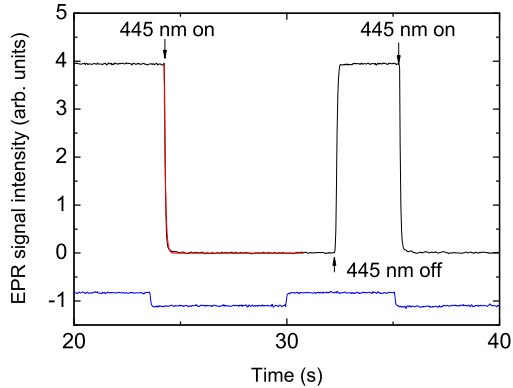}
\caption{\label{fig5} 
Kinetics of Mn$^{2+}$ photoquenching under 445~nm excitation at a power of 190 (upper trace) and 20~mW (lower trace). Red line is calculated for an exponential decay, $\exp(-t/\tau)$, with  $\tau=63$~ms.
}
\end{center}
\end{figure}
The kinetics of Mn$^{2+}$ photoionization is very abrupt, as shown in Fig.~\ref{fig5}. The kinetics was measured at 3~K at a constant field value, corresponding to a maximum intensity of one of the resonance lines. The upper and lower traces show the change of the signal intensity under 445~nm excitation at a power of 190 and 20~mW, respectively. The red line is calculated assuming a decay time of 63~ms 
(apparatus response time). 
This decay constant is much too short to account for the kinetics of a process involving charge transfer between Mn and other trap centers, which is usually of the order of minutes.~\cite{Godlewski1985} Moreover, the observed decay time does not depend on excitation power, which suggests that the real decay time is shorter than the spectrometer response. However, at wavelengths above 600~nm there is also a slower component, responsible for a few percent of the signal decrease.

The photo-EPR experiments have confirmed that the Mn$^{3+}$/Mn$^{2+}$ level is located at 2.1~eV below the conduction band minimum of ZnO. In addition, it was shown that in as grown ZnO: Mn crystals the manganese impurity occurs predominantly in the 3+ charge state. To account for the  partial occupancy of the Mn$^{2+}$ state, there have to be other acceptors centers in the sample which push the Fermi level below the impurity level. One of the candidates is the complex of manganese with interstitial oxygen, Mn-O$_i$, postulated by Gluba and Nickel.~\cite{Gluba} The presence of such an acceptor center, however, is not confirmed in our experiment. We observe only the recharging of isolated Mn ions. It should be also noted that apart from Mn$^{2+}$ we detect no other EPR signals, whether of acceptors nor donors, in our crystals. 

In particular, the EPR signal of a residual donor with the $g-$factor of 1.956, identified as hydrogen related shallow donor,~\cite{Detlev} is not observed even after hydrogenation of the sample. This signal does not appear also under illumination, which suggests that if this donor is present it is not effectively populated, {\it i.e.}, the electron capture rate is much lower than that of Mn$^{3+}$ ions. It should be also stressed that the temperature dependence of Mn$^{2+}$ photo-quenching is not governed by activation energies typical for ZnO donors, which range from 35 to about 70~meV.~\cite{Meyer} Instead, activation energies of the order of 1~meV, slightly dependent on sample treatment, are detected. All these point out to the conclusion that another temperature dependent mechanism must lead to a metastable change of the Mn$^{2+}$ occupancy under illumination.

%%%%%%%%%%%%%%%%%%%%%%%%%%%%%%%%%%%%%%%%%%%%%
\section{\label{sec3}Theory}

\subsection{\label{sec3a}Calculation details}
The calculations are performed within the density functional theory in the generalized gradient approximation (GGA) of the exchange-correlation potential.~\cite{Hohenberg,KohnSham,PBE} The $+U$ corrections are included.~\cite{Anisimov1991, Anisimov1993, Cococcioni} We use the pseudopotential method implemented in the QUANTUM ESPRESSO code,~\cite{QE} with the valence atomic configuration $3d^{10}4s^2$ for Zn, $2s^2p^4$ for O and $3s^2p^6 4s^2p^0 3d^5$ for Mn, respectively. 
The plane-waves kinetic energy cutoffs of 30~Ry for wavefunctions and 180~Ry for charge density are employed. The electronic structure of the wurtzite ZnO is examined with a $8\times 8\times 8$ $k$-point grid. Analysis of a single Mn impurity in ZnO is performed using $3\times 3\times 2$ supercells with 72 atoms (2.8 atomic per cent of Mn). $k$-space summations are performed with a $3\times 3\times 3$ $k$-point grid for density of states (DOS) calculations, 
while calculations with fixed occupation matrices 
are performed using the $\Gamma$ point only. The $U$ terms for $3d$(Zn), $2p$(O), and $3d$(Mn) orbitals are treated as free parameters, whose values are discussed below. Ionic positions are optimized until the forces acting on ions became smaller than 0.02~eV/\AA.

\subsection{\label{sec3b}Pure ZnO}
It was previously shown that both the local density approximation (LDA) and GGA fail to give correct band characteristics of ZnO. In particular, the band gap, $E_{gap}$, of ZnO calculated within LDA/GGA~\cite{Schroer, Jaffe, Lim} is about 1~eV. This is due to the universal "band gap problem", {\it i.e.}, the underestimation of the gap within LDA/GGA on the one hand, but also to the too high calculated energies of the $d$(Zn)-derived bands~\cite{Wei} on the other hand. The inclusion of the $U$(Zn) term~\cite{Zhou, Lim, Dong} solves this problem only partially, since the band gap is still underestimated by about 2~eV. For example, we find that when $U$(Zn)=10~eV is employed the $d$(Zn) band is at about 8~eV below the valence band maximum (VBM), in agreement with experiment,~\cite{Lim, Dong, Ley, Vesely} but $E_{gap}\approx 1.2$~eV is still wrong. 
This is because the coupling between $d$(Zn) and VBM is weak due to the large energy difference between those states, and thus $E_{gap}$ is not sensitive to the energy of the $d$(Zn) band. To obtain a correct value of $E_{gap}$ one should observe that the upper valence band is derived from $p$(O) orbitals. Indeed, the inclusion of the $U$(O) term for the $p$(O) orbitals, in addition to $U$(Zn), gives a correct band structure.~\cite{Ma, Lim, Agapito} We find that $U$(Zn)=12.5~eV and $U$(O)=6.25~eV reproduce both the experimental $E_{gap}$ of 3.3~eV~\cite{Dong} and the energy of the $d$(Zn) band, centered about 8~eV below the VBM, in excellent agreement with Ref.~\onlinecite{Agapito}. These values also lead to the correct width of $\sim 6$~eV of the upper valence band of mostly $2p$(O) character, and the lower conduction band of $4s$(Zn) character. 
The relaxed crystal structure agrees well with experiment: the lattice parameters $a = 3.23$~\AA\ and $c = 5.19$~\AA, as well as the internal parameter $u = 0.38$ are underestimated by less than 1~\% in comparison with experimental values: 
$a = 3.25$~\AA, $c = 5.20$~\AA, and $u = 0.38$.~\cite{Karzel} One should finally observe that the electronic structure of ZnO represents a problem even for the GW approach: as it is discussed in Refs.~\onlinecite{Lim, Lany} different  GW calculations, including quasiparticle self-consistent GW calculations, still place the $d$(Zn) band at an energy too high by about 1~eV, and an additional potential on Zn cations is needed to achieve the correct band structure.

\subsection{\label{sce3c}Mn impurity in ZnO}
\begin{figure}[t!]
\begin{center}
\includegraphics[width=8.3cm]{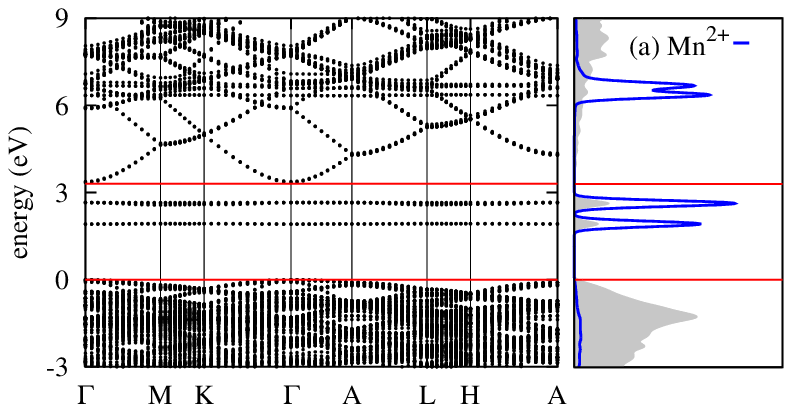}
\includegraphics[width=8.3cm]{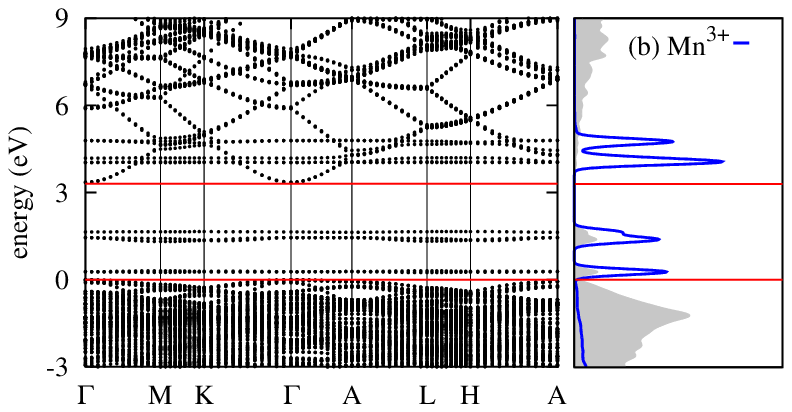}
\end{center}
\caption{\label{fig6} 
Energy bands (left panels) and DOS (right panels) of (a) ZnO:Mn$^{2+}$, and (b) ZnO:Mn$^{3+}$. Gray area and black lines (blue lines in the on line version) in DOS display the total DOS and DOS projected on $d$(Mn) orbitals, respectively. Horizontal (red) lines denote the band gap of ZnO.  
}
\end{figure}
Properties of the Mn ion in ZnO depend on its charge state. The band structure and DOS of ZnO doped with Mn$^{2+}$ and Mn$^{3+}$ are shown in Fig.~\ref{fig6} for $U$(Mn)=0. Mn$^{2+}$ introduces two levels into the gap, a $t_{2\uparrow}$ triplet at 2.64~eV above VBM and an $e_{2\uparrow}$ doublet at 1.90~eV. (Actually, $t_{2\uparrow}$ is split into a singlet and a doublet by the wurtzite crystal field with a small splitting of about 0.1~eV.) The spin-down states form resonances degenerate with the conduction band, and thus, in agreement with experiment, Mn cannot assume the 1+ charge state in n-type ZnO. 

The $t_{2\uparrow}$ and $e_{2\uparrow}$ levels of Mn$^{3+}$ are at about 1.45 and 0.28~eV above VBM, respectively, {\it i.e.}, they are lower by $\sim 1.5$~eV than those of Mn$^{2+}$. This large difference in the level energies of Mn$^{2+}$ and Mn$^{3+}$ stems from the strong intra-center Coulomb repulsion between $d$(Mn) electrons caused by the localization of their wavefunctions. Moreover, the localized character of $d$(Mn) is responsible for the relatively large 6~\% reduction of the Mn-O bond length, from 2.02~\AA\ for Mn$^{2+}$ to 1.90~\AA\ for Mn$^{3+}$, which is induced by the decrease in the Coulomb coupling between Mn and O anions. We also mention that the energies of the gap states of the isolated Mn$^{3+}$ and those of Mn$^{3+}$ with a photoelectron $e_{CB}$ in the conduction band are the same to within 0.02~eV, and the Mn-O bond lengths are the same to within 0.01~\AA. This is because of the delocalized character of the wave function from the bottom of the 
conduction 
band. The results for Mn$^{3+}$ with $e_{CB}$ we are shown in Fig.~\ref{fig8}c.

\subsection{\label{sec3d}Photoionization, recombination, and mechanism of metastability}
\begin{figure}[t!]
\begin{center}
\includegraphics[width=8.3cm]{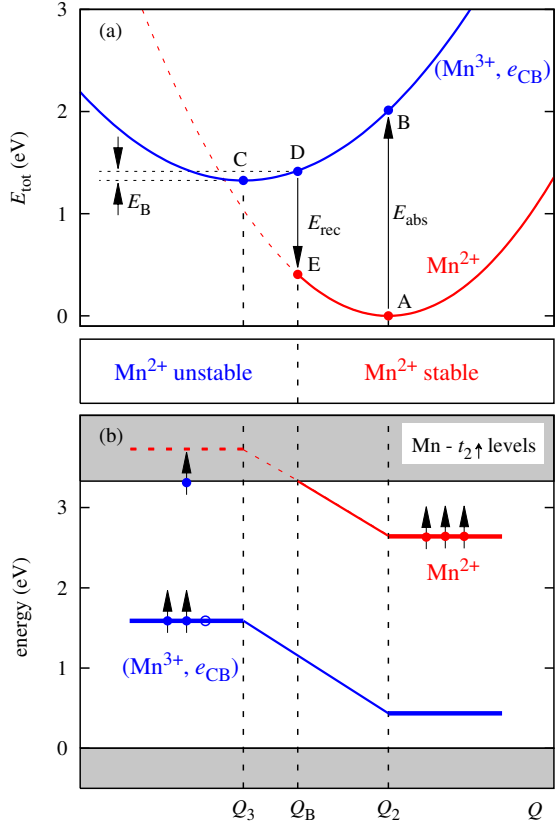}
\caption{\label{fig7} 
(a) Total energy change of Mn$^{2+}$ and (Mn$^{3+}, e_{CB}$) as a function of configuration coordinate $Q$. $Q_2$ and $Q_3$ are equilibrium atomic configurations of Mn$^{2+}$ and and (Mn$^{3+}, e_{CB}$) charge states, respectively, and $Q_B$ is the configuration coordinate of the barrier. (b) Single particle energy of the $t_{2\uparrow}$ level for both Mn$^{2+}$ (red symbols) and (Mn$^{3+}, e_{CB}$) (blue symbols); note the strong dependence of the $t_{2\uparrow}$ energy on the charge state. 
$U$(Mn)=0 is assumed. 
}
\end{center}
\end{figure}

\begin{figure*}[t!]
\begin{center}
\includegraphics[width=2.86cm]{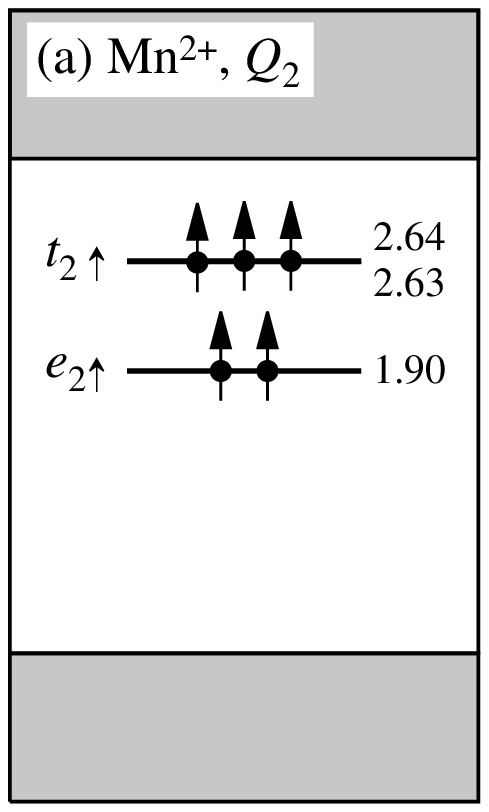}
\includegraphics[width=2.86cm]{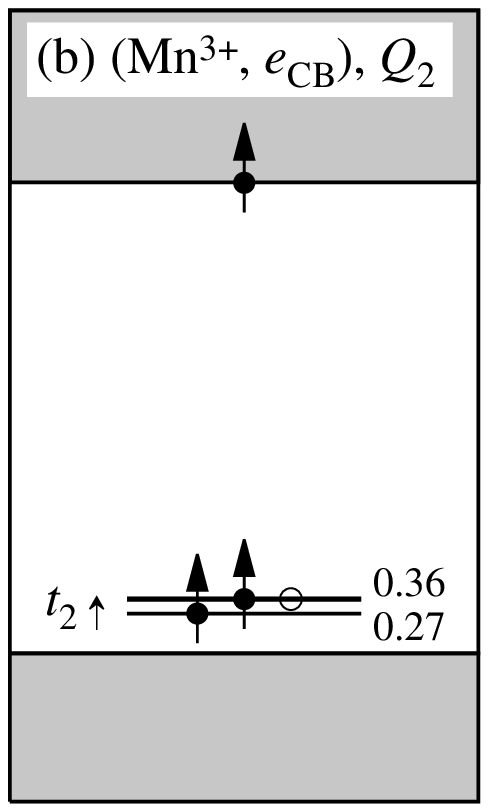}
\includegraphics[width=2.86cm]{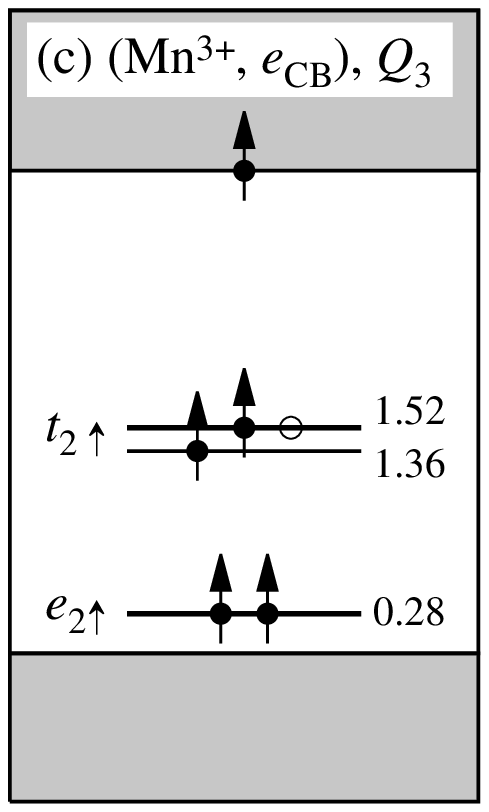}
\includegraphics[width=2.86cm]{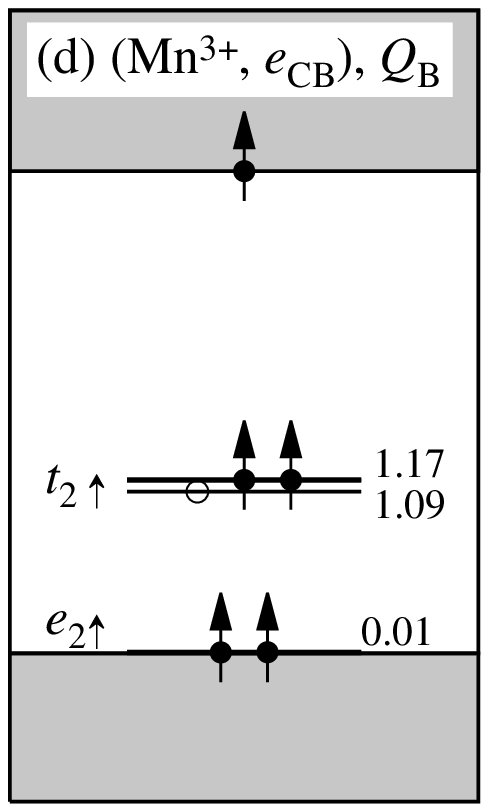}
\includegraphics[width=2.86cm]{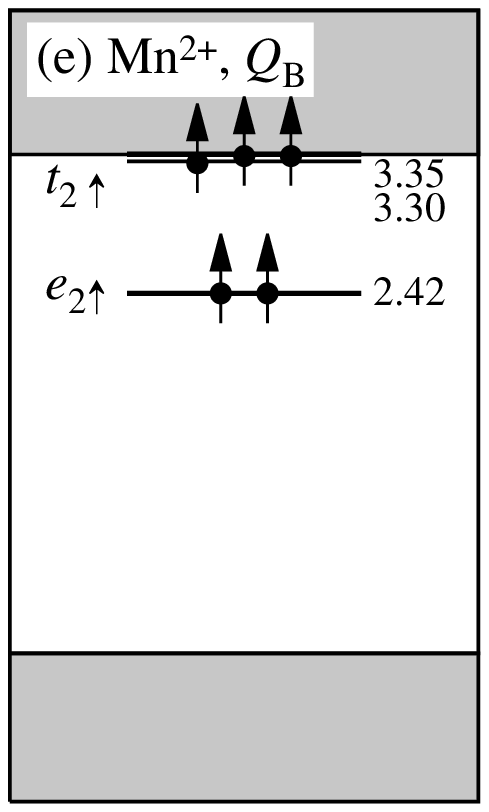}
\includegraphics[width=2.86cm]{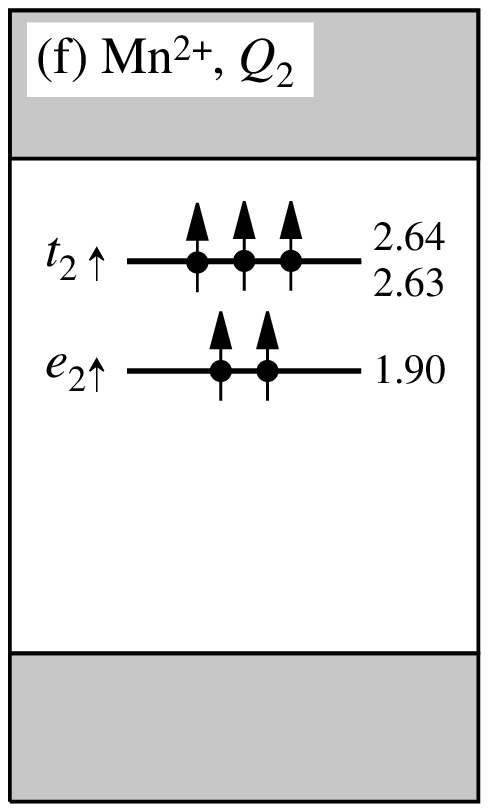}
\end{center}
\caption{\label{fig8} 
Calculated diagrams of Mn levels for  (a) Mn$^{2+}$ at equilibrium configuration $Q_2$, (b) Mn$^{3+}$ with a photoelectron $e_{CB}$ in the CB in the same atomic configuration $Q_2$, (c)  Mn$^{3+}$ with $e_{CB}$ in the relaxed configuration $Q_3$, (d) Mn$^{3+}$ with $e_{CB}$ in the barrier configuration $Q_B$, (e) Mn$^{2+}$ (when $e_{CB}$ is captured by Mn) in the barrier configuration $Q_B$, and (f) Mn$^{2+}$ relaxed to $Q_2$; this is the end step of the recombination process, the  configuration is the same as in (a). $U$(Mn)=0 is assumed. 
}
\end{figure*}
Due to the strong dependence of gap levels on the Mn charge state, the energies of absorption and/or recombination cannot be deduced directly from single particle states of Mn$^{2+}$ (or Mn$^{3+}$), as it was indicated in, {\it e.g.}, Refs.~\onlinecite{Badaeva, Gamelin}. Consequently, energies of processes analyzed below are calculated from the total energy difference between final and initial states.~\cite{note1}

The absorption-recombination cycle of Mn$^{2+}$ occurs in five steps. They are presented in Fig.~\ref{fig7}, which shows both the total energy and the Mn energy levels for each step for $U$(Mn)=0. Mn levels are shown in Fig.~\ref{fig8} in detail. 

(i) In the first step (Fig.~\ref{fig7}a, A~$\to$~B) one electron from $t_{2\uparrow}$ of Mn$^{2+}$ is excited to the conduction band, with the atomic positions kept fixed at the equilibrium configuration of Mn$^{2+}$, $Q_2$. The excitation energy is $E_{abs}=2.0$~eV. Photoionization induces a strong decrease of the $t_{2\uparrow}$ energy by about 2~eV, see Figs~\ref{fig7}b and \ref{fig8}b, because the depopulation of the $d$(Mn) shell reduces the strength of the Coulomb repulsion. 

(ii) In the second step (B~$\to$~C), atoms are allowed to relax towards the equilibrium configuration $Q_3$ of Mn$^{3+}$ with the photoelectron $e_{CB}$ in the conduction band. This case is denoted by (Mn$^{3+}, e_{CB}$) in Figs~\ref{fig7} and \ref{fig8}. During this step the Mn-O bonds are reduced by $\sim 6$~\%, and the energy of $t_{2\uparrow}$ increases by about 1~eV (Figs~\ref{fig7}b and \ref{fig8}c), in agreement with its antibonding character. The corresponding energy gain ($E_{tot}$(B)$- E_{tot}$(C)) is 0.69~eV.

This energy gain takes place in spite of the fact that the single particle gap level $t_{2\uparrow}$ increase in energy by more than 1~eV, see Figs.~\ref{fig7}b and \ref{fig8}c. This illustrates the fact that total energy differences cannot be deduced directly from single particle states of Mn$^{2+}$ (or Mn$^{3+}$), since other factors such as the Madelung ion-ion energy are dominant. 

According to our results, the relaxed (Mn$^{3+}, e_{CB}$) state of Mn$^{3+}$ with one electron in the conduction band is metastable, because its energy is higher than that of the relaxed Mn$^{2+}$ by 1.32~eV (($E_{tot}$(C)-$ E_{tot}$(A)) in Fig.~\ref{fig7}a), but a direct recombination of the photoelectron to the $t_{2\uparrow}$  level of Mn$^{2+}$ is not possible. The instability stems from the fact that in the configuration $Q_3$ the energy of the $t_{2\uparrow}$ level of Mn$^{2+}$ occupied with 3 electrons is above the conduction band bottom (CBB), see Fig.~\ref{fig7}b. Indeed, the calculated dependence of $t_{2\uparrow}$ of Mn$^{2+}$ on the configuration coordinates, presented in Fig.~\ref{fig7}b, shows that $t_{2\uparrow}$ increases in energy with the decreasing Mn-O bond lengths, and merges with the conduction band for the atomic configuration $Q_B$. 
For smaller bond lengths, in particular in the $Q_3$ configuration, it is a resonance degenerate with the conduction band. 
The extrapolated $t_{2\uparrow}$ energies are shown by dashed lines in Fig.~\ref{fig7}b, and the corresponding extrapolated total energy is shown by the dashed line in Fig.~\ref{fig7}a. We use extrapolated values because for the configuration coordinates in the range $(Q_B,Q_3)$ we could not arrive at convergent results when fixing the occupation of the $t_{2\uparrow}$ level by 3 electrons. 
The occupancy of $t_{2\uparrow}$ by 3 electrons is unstable since there are empty conduction states lower in energy. In other words, Mn$^{2+}$ is stable for configuration coordinates in the range from $Q_2$ to $Q_B$, when $t_{2\uparrow}$ is a gap state, while (Mn$^{3+}, e_{CB}$) is locally stabilized for configuration coordinates in the range ($Q_3$, $Q_B$). At $Q_B$, the electron recombination is possible, with the corresponding energy gain $E_{rec}$. The difference in total energy of (Mn$^{3+}, e_{CB}$) between $Q_3$ and $Q_B$ is the energy barrier $E_B$. 

The difference between total energy of Mn$^{2+}$ and (Mn$^{3+}, e_{CB}$) states in the $Q_3$ configuration is only estimated from single particle levels at the state halfway between, thanks to the Janak theorem.~\cite{Janak} It formally gives the lower energy of Mn$^{2+}$ state by 0.59~eV. Actually, however, the Mn$^{2+}$ state is unreachable in the $Q_3$ configuration and does not converge in our calculations. The Janak theorem can be applied in this case, since the ionic positions are kept fixed at $Q_3$. 

The recombination of $e_{CB}$ requires three more steps shown in Figs~\ref{fig7} and \ref{fig8}. 

(iii) A thermally driven atomic transition from $Q_3$ to the barrier configuration $Q_B$ (C~$\to$~D), which is described in detail below. According to our estimates, the upper limit for the barrier $E_B$ is 60~meV. 

(iv) The capture of the photoelectron by Mn$^{3+}$ (D~$\to$~E), {\it i.e.}, the transition to the Mn$^{2+}$ state. This transition at the estimated $Q_B$ configuration provides the energy gain  of $E_{rec} = 0.98$~eV.

(v) The relaxation of Mn$^{2+}$ from $Q_B$ to $Q_2$ (E~$\to$~A) with $E_{relax}= 0.41$~eV. 

Finally, we notice that the metastable atomic configuration $Q_3$ of (Mn$^{3+}$, $e_{CB}$) is an excited state of the crystal as a whole, since the corresponding total energy  is higher than that of ZnO:Mn$^{2+}$ in the ground state configuration $Q_2$. 
Importantly, however, in those particular atomic configurations electrons are in the respective ground states, which justifies the usage of GGA$+U$. 
Consequently, both the total crystal energies and the total energy difference between the metastable configuration $Q_3$ and the ground state $Q_2$ are well defined as well. 
Moreover, in the $Q_3$ configuration small displacements of anions around Mn increase the total energy of (Mn$^{3+}, e_{CB}$), which proves that this is indeed a metastable state of the crystal, and the barrier for electron recombination is non-vanishing.

\subsection{\label{sec3e}Estimation of the energy barrier}
\begin{figure}[t!]
\begin{center}
(a)\includegraphics[width=3.5cm]{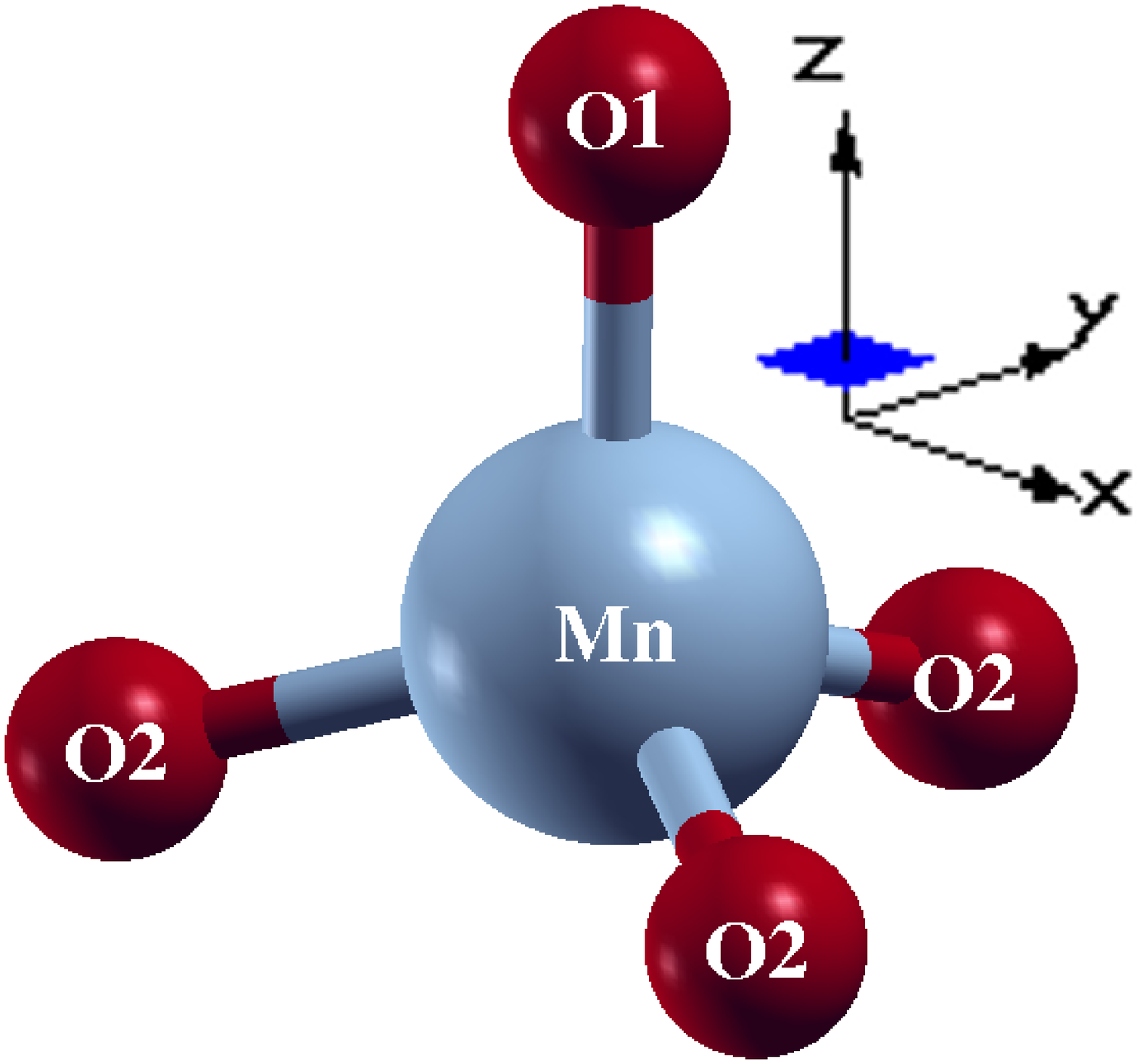}
(b)\includegraphics[width=4.0cm]{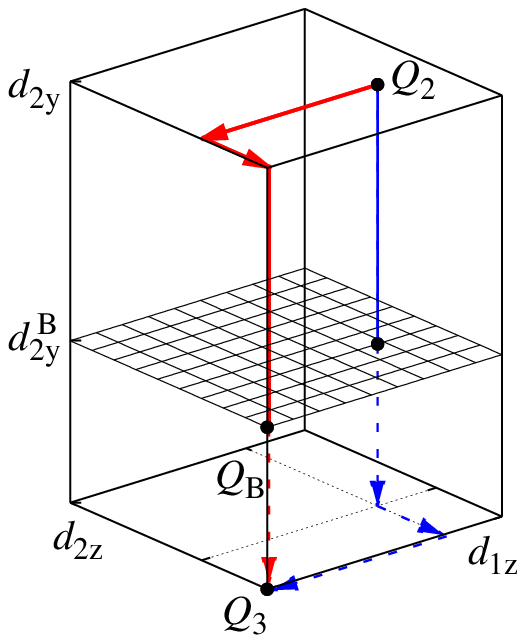}
\caption{\label{fig9} 
(a) Mn-O bonds. The atomic positions in cartesian coordinates are Mn$(0,0,0)$, O1$(0,0,d_{1z})$, O2$(0,d_{2y} ,d_{2z})$. (b) Two path between $Q_2$ and $Q_3$. For $d_{2y}^B$, Mn$^{2+}$ becomes unstable, {\it i.e.}, the $t_{2\uparrow}$ level of Mn$^{2+}$ is above CBM. $Q_B$ is the estimated barrier configuration.
}
\end{center}
\end{figure}

\begin{table}
\caption{\label{tabI}
Equilibrium coordinates of O1 and O2 shown in Fig.~\ref{fig9}a, the respective Zn-O and Mn-O bond lengths 
$d_1$ and $d_2$, and the average bond length $< d >= (d_1 + 3d_2)/4$. All values are in \AA.
}
\begin{ruledtabular}
\begin{tabular}{l c c c c c}
&	$d_1=d_{1z}$ & $d_{2y}$ & $d_{2z}$ & $d_2$ & $<d>$\\ 
\hline
ZnO:&    	1.98& 1.87& -0.62& 1.97& 1.97\\
Mn$^{2+}$:& 2.03& 1.92& -0.63& 2.02& 2.02\\
Mn$^{3+}$:& 1.94& 1.79& -0.59& 1.88& 1.90\\
\end{tabular}
\end{ruledtabular}
\end{table}
A detailed description of the metastability, in particular of the barrier height for return to the ground state, is difficult, because the atomic relaxations around Mn involve not only the nearest but also more distant neighbors. To make the problem tractable, we limit the parameter space to the four Mn-O bonds shown in Fig.~\ref{fig9}a. The local symmetry of Mn is $C_{3v}$ in all the considered cases, and consequently there are 3 parameters that define the geometry, $d_{1z}$, $d_{2y}$, and $d_{2z}$, which are defined in the caption to Fig.~\ref{fig9}. The three basal O atoms are equivalent.  Mn is assumed to be at $r=0$, and the atoms beyond the first neighbors are allowed to relax. 
The calculated coordinates of the two non-equivalent oxygen ions for both Mn$^{2+}$ and Mn$^{3+}$ in the (Mn$^{3+}, e_{CB}$) configuration, together with the Zn-O bond lengths in ZnO for comparison, are given in Table~\ref{tabI}. 
Two possible paths between $Q_2$ and $Q_3$ are displayed in Fig.~\ref{fig9}b. In both cases we found that the $t_{2\uparrow}$ level of Mn$^{2+}$ is much more sensitive to the changes of $d_{2y}$ than of $d_{1z}$ or $d_{2z}$. 
For both paths, the Mn$^{2+}$ instability begins at almost 
the same $d_{2y}$, which is denoted by $d_{2y}^B$ in Fig.~\ref{fig9}b.  Therefore, the barrier configuration $Q_B$ is taken as a point which is achieved from $Q_3$ by changing only the $d_{2y}$ coordinate. 
With this assumption we find $Q_B$ as the configuration at which the $t_{2\uparrow}$ level of Mn$^{2+}$ is degenerate with CBM. This allows us to find the corresponding energy barrier, $E_B=E($Mn$^{3+},Q_3)-E($Mn$^{3+},Q_B) = 60$~meV, which clearly represents the upper limit. 

\subsection{\label{sec3f} Dependence on $U$(Mn)}
\begin{figure}[t!]
\begin{center}
\includegraphics[width=8.3cm]{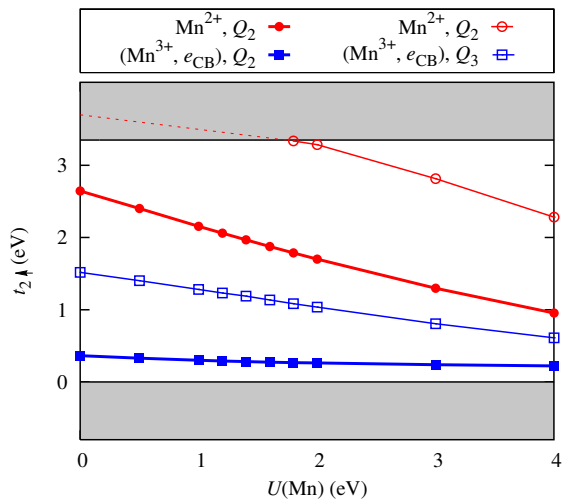}
\caption{\label{fig10} 
Dependence of the $t_{2\uparrow}$ energy level on $U$(Mn)
for 
Mn$^{2+}$ and (Mn$^{3+},e_{CB}$) in the configurations $Q_2$ and $Q_3$. 
}
\end{center}
\end{figure}

\begin{figure}[t!]
\begin{center}
\includegraphics[width=8.3cm]{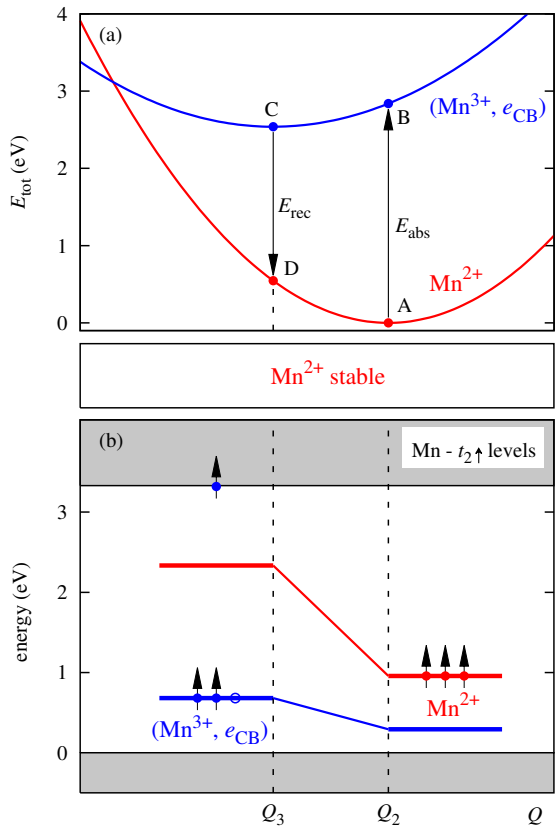}
\caption{\label{fig11} 
The $Q$ dependence of (a) the total energy and (b) single particle levels of Mn$^{2+}$ and (Mn$^{3+}, e_{CB}$) for $U$(Mn) = 4~eV. 
}
\end{center}
\end{figure}

The analysis presented in the previous Section was conducted assuming $U$(Mn)=0. As it was mentioned in Sec.~\ref{sec3a}, the value of $U$(Mn) is treated here as a free parameter, which can be adjusted to fit the experimental data. We have performed calculations for a few values of $U$(Mn), and the results are presented in Fig.~\ref{fig10}.
As is follows from Fig.~\ref{fig10}, the energies of gap levels of both  Mn$^{2+}$ and Mn$^{3+}$ decrease with increasing $U$. 
In particular, assuming $U$(Mn)=4~eV brings $t_{2\uparrow}$ about 0.9~eV above the VBM, and puts the $e_{2\uparrow}$ level below the VBM. 
We also note that for $U$(Mn)=4~eV, in the equilibrium configuration $Q_2$ Mn-O bond lengths are 2.06~\AA, while they are reduced to about 1.98~\AA\ for $Q_3$, which shows that the impact of $U$ on bond lengths is moderate.

Comparing our results with previous theoretical investigations of Mn in ZnO
we note that the LDA calculations including SIC corrections~\cite{Toyoda} were performed for high Mn content, for which a wide Mn-induced band in the band gap was found in qualitative agreement with our results.
LDA supplemented with the $+U$ term imposed on the $d$(Mn) orbitals was also used~\cite{Raebiger, Chanier, Gluba}. 
For $U$(Mn)=3~eV, there is a reasonable agreement with Ref.~\onlinecite{Raebiger}, which uses $U=3.9$~eV and $J=1$~eV (this corresponds to the effective $U=3.9-1.0=2.9$~eV), and with the time dependent DFT.~\cite{Badaeva}
The other applied $U$(Mn) values were 6~eV~\cite{Chanier} obtained from the fit to the experimental magnetisation data, and 3.2~eV~\cite{Gluba} estimated according to Ref.~\onlinecite{Janotti}. 
In these works, the Mn$^{2+}$ $t_{2\uparrow}$ level is situated at about 0.7-1.0~eV above the VBM. 
The levels of Mn$^{3+}$ were not investigated. 
Our results obtained with $U$(Mn)=3-4~eV are reasonably close to those quoted above.  

To obtain the optimal value of $U$(Mn) by fitting to our experimental results we note that 
the $U$-induced downward shifts of the Mn levels imply that the excitation energy (step A~$\to$~B in Fig.~\ref{fig7}a) increases from 2.0 to 2.84~eV when $U$ changes from 0 to 4~eV. 
Moreover, the barrier $E_B$ depends on the $U$(Mn) term: it decreases with the increasing $U$ and it vanishes for $U$(Mn) higher than about 1.7~eV. The decrease of $E_B$ is related with the $U$-induced decrease of the Mn levels. In particular, for $U=0$ the $t_{2\uparrow}$ level is degenerate with the conduction band for the configuration $Q_3$, while for $U>1.7$~eV it is below the CBM, and therefore a direct transition of the photoelectron from the conduction band to $t_{2\uparrow}$ is possible. This feature is illustrated in Fig.~\ref{fig11} for $U$(Mn) = 4~eV. 
Therefore, the best overall value of $U$(Mn) is about 1.5~eV, giving a barrier of about 1~meV, and the excitation energy of about 2.4~eV. 

%%%%%%%%%%%%%%%%%%%%%%%%%%%%%%%%%%%%%%%%%%%%%
\section{\label{sec4}Summary}
Photo-EPR experiments performed on ZnO:Mn single crystals have confirmed that the Mn$^{3+}$/Mn$^{2+}$ level is located about 2.1~eV below the conduction band minimum of ZnO, as illumination of the crystal with photon energies higher than 2.1~eV leads to a partial, temperature dependent depopulation of the Mn$^{2+}$ state at low temperatures, accompanied by photoconductivity. The unusually small thermal deactivation energy (of the order of 1~meV) together with the untypically fast kinetics of Mn$^{2+}$ photoquenching point out to a process different from charge transfer from Mn$^{2+}$ to other defect centers. We interpret the observed metastable change of Mn$^{2+}$ occupancy under illumination as due to a small energy barrier for electron recapture from the conduction band by Mn$^{3+}$.  

GGA+$U$ approach was employed to study both the Mn$^{2+}$ - Mn$^{3+}$  
optical transitions, and the stability of the (Mn$^{3+}$, $e_{CB}$) photoexcited state.  
The excited state  is found to be metastable, because in the relaxed configuration a direct recombination of the photoelectron is not possible, and recapture requires overcoming an energy barrier. 
The energy barrier $E_B$ decreases with the incrasing $U$(Mn), and vanishes for $U>1.7$~eV. 
Comparing theory with experiment we find that $U$(Mn) of about 1.5~eV leads to photoionization energy, 2.4~eV, and the barrier $E_B$ of the order 1 meV, in a reasonable  agreement with the experimental data. 
Moreover, one should note that similar results hold for Mn and Fe ions in GaN,~\cite{VZB, ZB} for which the experimental intra-center transition energies are reproduced with very small $U$ terms. 

The metastability of (Mn$^{3+}$, $e_{CB}$) is related with the reduction of the Mn-O bonds after the Mn$^{2+}$ ionization. 
This change of atomic configuration is coupled to electronic degrees of freedom, 
and it rises the energy of the d(Mn) donor level. 
More importantly, the Coulomb repulsion between the $d$(Mn) electrons is strong, 
and it rises the Mn level energy by $\sim 1$~eV when Mn changes its charge state to Mn$^{2+}$. 
In more detail, while the donor level of Mn$^{3+}$  is situated below the bottom of the conduction band, 
after capturing the photoelectron the occupied donor level of Mn$^{2+}$  would be 
above the empty conduction band, which is an unstable electronic configuration. 
This factor blocks the recombination of the photoelectron, and drives the metastability of Mn$^{3+}$. 
While the role of the local lattice relaxatons  was recognized and extensively discussed 
for metastable centers in semiconductors, 
~\cite{EL2, EL2_prl, Chadi, Dobaczewski, BBdx, Wetzel, Thio, Lany_DX, Jones, Schmidt}
the role of the strong Coulomb coupling between $d$ electrons represents a novel aspect of the physics of defect metastbility.

%%%%%%%%%%%%%%%%%%%%%%%%%%%%%%%%%%%%%%%%%%%%%
\section*{Acknowledgments}
The authors acknowledge the support from the projects No. 2012/05/B/ST3/03095 and 2011/01/D/ST7/02657, which are financed by Polish National Science Centre (NCN). Calculations were performed on ICM supercomputers of University of Warsaw (Grant Nos. G46-13 and G16-11).


\begin{thebibliography}{54}
\bibitem[{not({\natexlab{b}})}]{note1}
\bibinfo{note}{We have also checked that the integration of Mn level energies
  over fractional occupancies in the range between 0 and 1 reproduces the exact
  results for absorption, in accordance with Janak's theorem~\cite{Janak},
  while the approximation by the midpoint value (occupancy 1/2) gives an error
  lower than 0.2~eV.}
\end{thebibliography}
\end{document}